\pdfoutput = 1
\documentclass[
    reprint, 
    amsmath,
    amssymb,
    aip,
    apl,
    superscriptaddress,
    showkeys,
    floatfix
]{revtex4-2}

\usepackage{graphicx} 
\usepackage{dcolumn}
\usepackage{bm}
\usepackage{multirow}
\usepackage{float}
\usepackage{textgreek}
\usepackage[version=3]{mhchem}
\usepackage{xcolor}
\usepackage{hyperref}
\usepackage[mathlines, pagewise]{lineno}
\usepackage[american]{babel}
\usepackage{url}
\usepackage{soul}
\usepackage{upgreek}

\definecolor{MZ}{RGB}{0, 0, 0}\def\MZ{\color{MZ}}

\definecolor{JL}{rgb}{0.0,0.0,0.0}

\definecolor{red}{rgb}{0,0,0}\def\red{\color{red}}

\begin{document}
\preprint{AIP/123-QED}

\title{
    {\red Crater-shaped Enrichment of $\mathrm{V}_\mathrm{Si}$ Color Centers in $4H$-SiC using Single-Pulse Near-Infrared Femtosecond Laser Processing}
}

\author{Mengzhi Yan}
\affiliation{State Key Laboratory of Precision Measuring Technology and Instruments, Laboratory of Micro/Nano Manufacturing Technology, Tianjin University, Tianjin 300072, China}

\author{Junlei Zhao} 
\email{zhaojl@sustech.edu.cn}
\affiliation{Department of Electronic and Electrical Engineering, Southern University of Science and Technology, Shenzhen 518055, China}

\author{Ying Song}
\affiliation{State Key Laboratory of Precision Measuring Technology and Instruments, Laboratory of Micro/Nano Manufacturing Technology, Tianjin University, Tianjin 300072, China}
 
\author{Bing Dong}
\affiliation{State Key Laboratory of Precision Measuring Technology and Instruments, Laboratory of Micro/Nano Manufacturing Technology, Tianjin University, Tianjin 300072, China}
 
\author{Yifei Duan}
\affiliation{State Key Laboratory of Precision Measuring Technology and Instruments, Laboratory of Micro/Nano Manufacturing Technology, Tianjin University, Tianjin 300072, China}
 
\author{Jianshi Wang}
\affiliation{State Key Laboratory of Precision Measuring Technology and Instruments, Laboratory of Micro/Nano Manufacturing Technology, Tianjin University, Tianjin 300072, China}
 
\author{Qingqing Sun}
\affiliation{State Key Laboratory of Precision Measuring Technology and Instruments, Laboratory of Micro/Nano Manufacturing Technology, Tianjin University, Tianjin 300072, China}
 
\author{Zongwei Xu}
\email{zongweixu@tju.edu.cn}
\affiliation{State Key Laboratory of Precision Measuring Technology and Instruments, Laboratory of Micro/Nano Manufacturing Technology, Tianjin University, Tianjin 300072, China}
 
\begin{abstract}
Currently, Si vacancy ($\mathrm{V}_\mathrm{Si}$) color centers in SiC are of significant interest due to their potential applications in quantum sensing and quantum communication.
Meanwhile, 
the qualities of laser-induced color centers are well guaranteed.
Femtosecond laser processing suffices for increasing the yield of $\mathrm{V}_\mathrm{Si}$ color centers in bulk materials and forms crater-shaped enriched regions on the surface.
However, 
there is a notable absence of existing simulation methods to explain the mechanisms behind laser-assisted $\mathrm{V}_\mathrm{Si}$ color center generation.
In this work, 
we design a three-dimensional molecular dynamics (3D-MD) model using an integral hemi-ellipsoidal shell mathematical model to simulate the interaction of Gaussian laser beams with bulk materials.
Furthermore, we calculate the transmittance, 
absorption coefficient, 
refractive index, 
and reflectivity of $4H$-SiC. 
Then, 
the absorptance of a 1030 nm laser in 350 $\upmu$m-thick $4H$-SiC material is abtained to simulate the energy loss during the actual processing. 
Finally, 
the study analyzes the movement trajectories of $\mathrm{V}_\mathrm{Si}$ color centers and explains the source of $\mathrm{V}_\mathrm{Si}$ on the surface.
This analysis explains the reasons for the enrichment of color centers in the crater-shaped regions formed after laser deposition.
Our work provides an effective 3D-MD modeling approach to study the processing mechanisms of laser interaction with semiconductor materials, 
offering insights into efficient $\mathrm{V}_\mathrm{Si}$ color center creation processes.

\end{abstract}

\maketitle


Silicon carbide (SiC) has good electrical and optical properties and is expected to replace silicon (Si) in the field of high-voltage and high-frequency power electronics~\cite{2019_property_SiC, 2011_elec_property, 2022_absorption_ntype}.
Furthermore, in quantum sensing area, color centers, known as flourescent point defects, is significant and deserved to be careful investigated~\cite{2020_quantum_review, 2016_review_quantum, 2014_quantum}.
Among numerous color centers in SiC, the silicon monovacancy ($\mathrm{V}_\mathrm{Si}$) is widely used in particular devices~\cite{2024_4H_colorcen},
for instance, the electrical devices~\cite{2018_elec_SiC}, 
magnetic devices~\cite{2023_SiC_Magnet}, and temperature sensing devices~\cite{2016_SiC_Therm, 2017_SiC_Therm}.
In order to efficiently generate the color center, ion implantation~\cite{2011_colorcen, 2014_ion_im}, chemical vapor growth~\cite{2014_chem_colorcen}, heat treatment~\cite{2015_ther_colorcen}, 
and femtosecond laser processing are always applied~\cite{2022_laser_SiC, 2016_review_quantum}.
Femtosecond lasers, with their low heat loss effect, offer advantages in preparing color centers in bulk materials.
Compared to the common continuous-wave laser (CW laser), 
femtosecond laser processing is tentatively used in SiC manufacturing~\cite{2016_SiC_laser_app},
for its higher absorption, and higher precision during irradiating materials~\cite{2023_4H_laserMD}.
Based on these properties, the femtosecond laser has capabilities to be applied in processing microstructures~\cite{2023_4H_laser} and generating useful defects~\cite{2011_colorcen, 2016_colorcen} for research needs.

However, 
the theory of femtosecond laser processing 
of semiconductor remains incomplete.
Finite element method (FEM) is still limited by the atomic scale information of the processed system~\cite{2023_4H_laserMD, 2016_SiC_COMSOL} and
Two-temperature model molecular dynamics (TTM-MD) is impossible to reflect the particle distribution and morphology information of 3D structure.~\cite{2024_SiC_TTMMD, 2023_SiC_TTMMD_An}.
The influence of {\red Gaussian-profile} laser beams on the surface morphology of materials and the distribution of color centers cannot be adequately represented.

In this work, 
considering the transmittance, reflectivity and absorption coefficient of $4H$-SiC to laser, 
a 20-layer, 3D {\red hemi-ellipsoidal} shells with energy distribution is designed to refine the Gaussian laser energy distribution.
Creating crater-like~\cite{2022_Cater_laser} processing marks on the surface of $4H$-SiC materials, successfully. 
For experiments,
{\MZ photoluminescence mapping (PL)} and laser confocal microscopy are conducted to evaluate the distribution of $\mathrm{V}_\mathrm{Si}$ and the surface topography of the processing areas.
The distribution of $\mathrm{V}_\mathrm{Si}$ after single-pulse processing is simulated by MD. 

Nitrogen-doped 350 ($\pm 50$) $\upmu$m-thick $4H$-SiC with {\MZ an epilayer} of 12.67 $\upmu$m is implanted with 270 keV $^{1}$H$^{+}$ ion with a dose of $1 \times 10^{16}$~cm$^{-2}$ to generate $\mathrm{V}_\mathrm{Si}$.
On this crystalline material, a near-infrared femtosecond laser (generated by HR-Femto-IR-10-10) with a wavelength of 1030 nm, a pulse width of 285 fs, a pulse repetition rate of 100 kHz,
a peak power ($Q_\mathrm{m}$) of $1.40 \times 10^{14}$~W$\cdot$cm$^{-2}$ and a waist radius of 2 $\upmu$m, is applied.
During the experiment, 
single point laser processing and array processing is carried out. 
The discussion of experimental and simulated results in the following parts are obtained from laser processing with pulse energy of 4000~nJ, while the rest of the results with lower pulse energy are shown in supplementary information (SI). 

In order to theoretically explain the phenomena of the color centers on the surface of the block after processing and correspond the actual processing conditions with the simulation results, 
the parameters consistent with the experiment are adopted as far as possible in the simulation process. 
Thereupon,
the simulation is carried out in two aspects: 
(i) the theoretical value of the energy absorbed by the Gaussian laser;
(ii) the MD simulation by extending the Gaussian energy to the ellipsoidal shell.

For the first aspect, 
reflectivity ($R$) for $4H$-SiC is generated from refractive indices.
The measured ordinary ($n_\mathrm{o}$) refractive index is fitted to the Sellmeier equation by using the least squares method~\cite{2013_4H_reflective}: 
\begin{equation}
\begin{aligned} \label{eq:refractive_n}
    n^{2}_{\mathrm{o-4H}} = & 1+\frac{0.20075\lambda^2}{\lambda^2+12.07224} + \frac{5.54961\lambda^2}{\lambda^2-0.02641}+ \\
    & \frac{35.65066\lambda^2}{\lambda^2-1268.24708},
\end{aligned}
\end{equation}
where the wavelength, $\lambda$, is in unit of micrometer. 
On the real domain, the reflectivity, $R$, is generated through the formula~\cite{2024_datSiC, 2007_reflect, 2013_Refection}:
\begin{equation} \label{eq:Reflectance_R}
    R = \frac{(n-1)^2+k^2}{(n+1)^2+k^2},
\end{equation}
where $n$ is ordinary refractive and $k$ is extinction coefficient.
$\alpha$ is treated as an absorption coefficient and calculated as 5.6282~cm$^{-1}$ for a 1030 nm laser.
Since the $4H$-SiC sample used in the experiment is $n$-type semiconductor material formed by nitrogen-doping, free electrons are introduced. The interaction of electrons with photons leads to enhanced absorption in the infrared region. 
Moreover, the absorption coefficient obtained by theoretical calculation is almost consistent with the test value provided in several experimental studies~\cite{2022_absorption_ntype, 2018_review_absorption, 1989_ori, 2001_absorb_SiC, 2016_SiC_COMSOL}: 
\begin{equation} \label{eq:absorption_a}
    \alpha = \frac{4 \pi k}{\lambda},
\end{equation}
Transmittance ($T$) for $4H$-SiC is defined as~\cite{2002_optics_ART}: 
\begin{equation} \label{eq:Transmittance_T}
    T = (1-R)^{2}\exp{(- \alpha d)},
\end{equation}
where $R$ is reflectivity of $4H$-SiC and $\alpha$ is the optical absorption coefficient. 
Eq.~\ref{eq:Transmittance_T} is used to evaluate the value of $\alpha d$ using the measured transmission data~\cite{2024_coefficient_opt}. 
Through the actual results of the calculation,
transmittance value is significantly large. 
So in the absorption equation of energy, 
we ignore transmission intensity of 1030 nm laser in $4H$-SiC.
Therefore, the absorptance, $A$, is calculated from $R$ and $T$~\cite{2002_optics_ART} as: 
\begin{equation} \label{eq:Absorption_A}
    A = 1 - R - T,
\end{equation}
where the value of $A$ is $1.58\times10^{-3}$ in this work, as for $4H$-SiC, most of 1030 nm laser energy is reflected or transmitted, only little part of it is absorbed. 
Therefore, the effective processing energy, $E_\mathrm{eff}$, can be calculated from the incident energy, $E_\mathrm{incident}$), and the absorptance as:
\begin{equation} \label{eq:Absorption_A}
    E_\mathrm{eff} = E_\mathrm{incident} \cdot A.
\end{equation}
The {\MZ beam size} is applied as the actual processing energy to distribute the energy of the multi-layer simulated hemi-ellipsoidal shells to meet the physical absorption process of laser interaction with $4H$-SiC block.

\begin{figure}[ht!] 
    \includegraphics[width=8cm]{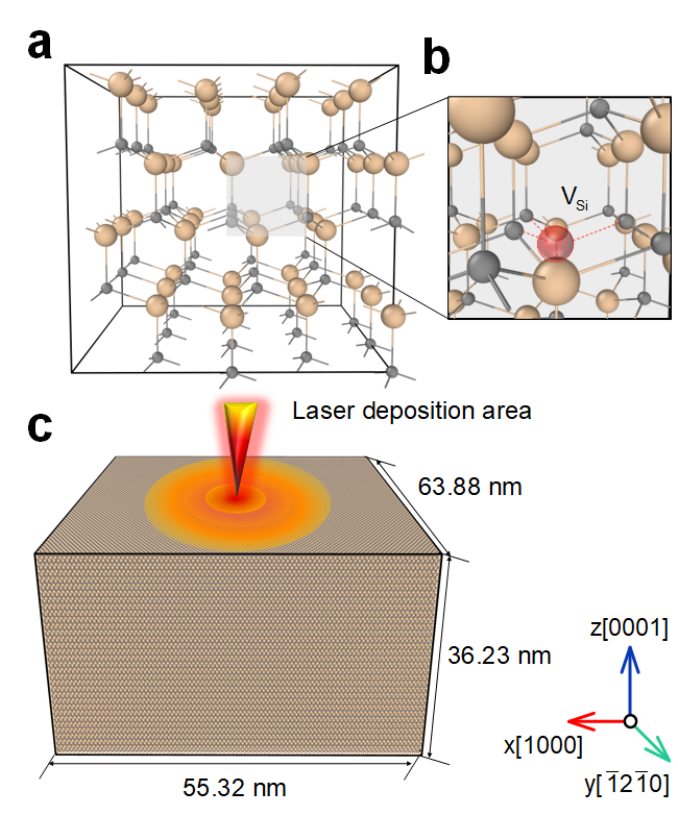}
    \caption{
        Simulation of $4H$-SiC system, the structure information of $\mathrm{V}_\mathrm{Si}$ and the deposition mode of laser energy.
        (a) The 95-atom $4H$-SiC cell, as the original object, is expanded into the structure shown in the panel (c).
        (b) The red transparent sphere represents the $\mathrm{V}_\mathrm{Si}$, and the red dotted line represents the Si-C bond with the first-nearest-neighboring C atoms of the $\mathrm{V}_\mathrm{Si}$. 
        (c) The laser is deposited from [0001] direction, 
        and the {\red beam size} meets the energy density distribution of Gaussian beam.
    }
    \label{fig:4H-SiC_VSi}
\end{figure}

Due to the computational cost, the size of the MD model cannot reach the scale of the actual bulk material, 
since the actual Gaussian beam waist radius is 2 $\upmu$m.
The final deposited energy needs to be reduced according to the radius of hemi-ellipsoidal shell and actual Gaussian beam waist, 
from the radius of 5 $\upmu$m to 10 nm spot. 
So the energy has to shrink by $1.25 \times 10^8$. 
{\MZ The basic structure of $4H$-SiC and $\mathrm{V}_\mathrm{Si}$ are shown in Figs.~\ref{fig:4H-SiC_VSi} (a), (b).}
{\red The actual MD simulation cell is illustrated in Fig.~\ref{fig:4H-SiC_VSi}(c).}

\begin{figure*}[htbp]
    \includegraphics[width=16cm]{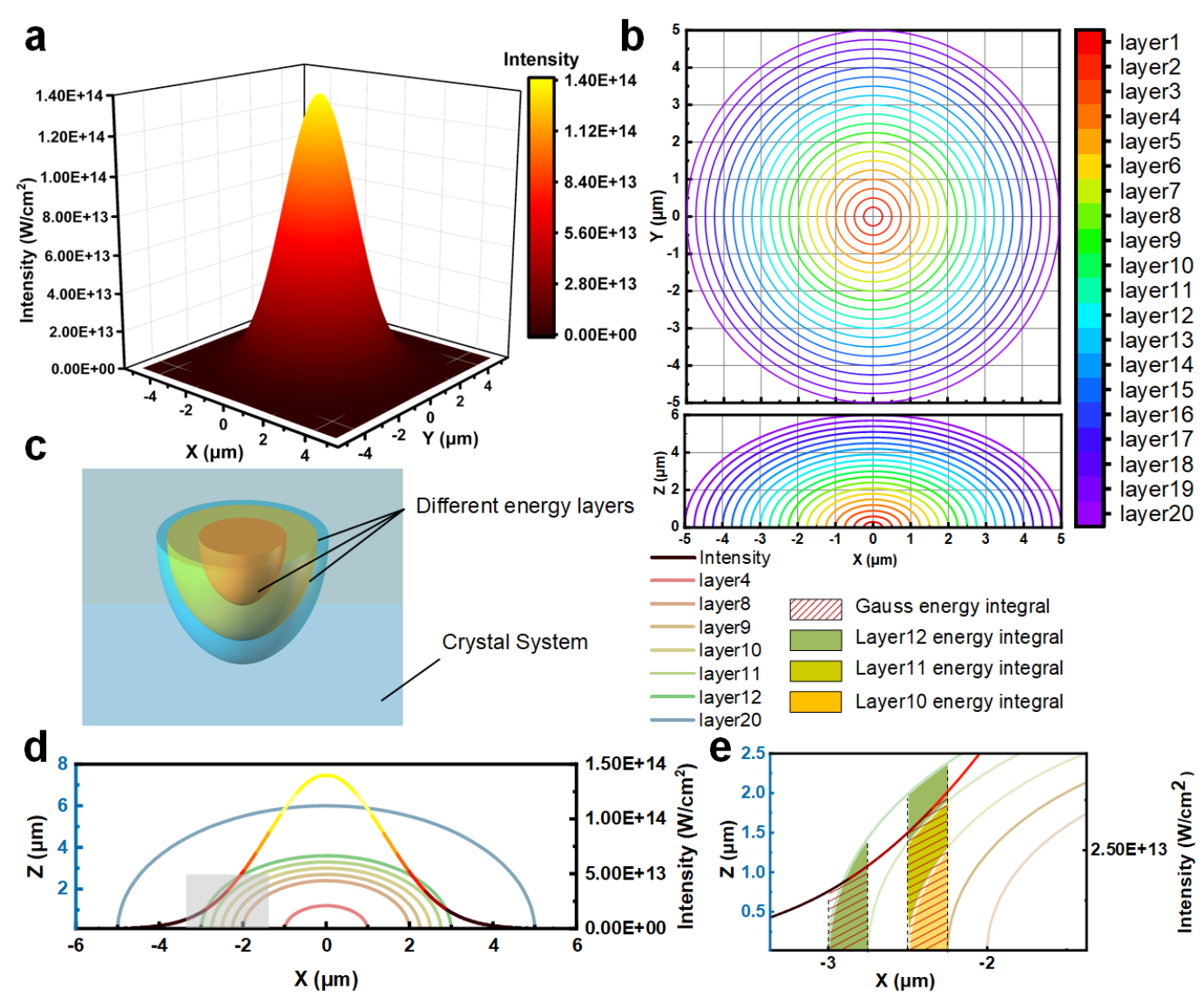}
    \caption {
    The integral relationship between the Gaussian energy and the corresponding energy in the hemi-ellipsoidal shell.
    (a) The maximum peak power of the laser used in the experiment is $1.4 \times 10^{14}$~W$\cdot$cm$^{-2}$, 
    and the beam waist radius is 2 $\upmu$m.
    (b) 20 layers are set from the inside to the outside. 
    (c) 3D-model of energy distribution upon the hemi-ellipsoidal shells and the block.
    (d) The cross sections of hemi-ellipsoidal shell and Gaussian energy are analyzed. 
    (e) The energy of 3D hemi-ellipsoidal shell and 3D Gaussian light is consistent in the corresponding region by means of volume division.}
    \label{fig:simulation_graph}
\end{figure*}

For second aspect,
as shown in Figs.~\ref{fig:simulation_graph} (a), (b), and (c), the energy of Gaussian laser is transition to the hemi-ellipsoidal shell model.
In order to ensure the uniformity of energy refinement, 
the number of hemi-ellipsoidal shells is set to 20 layers. 
For the energy distribution of each layer of particles, 
the data used in this paper is shared in SI.
During femtosecond laser processing, 
the absorption of energy is transformed from linear absorption to nonlinear absorption~\cite{2021_fslaser_theory},
so the circumference of model ellipsoid shell obtained by processing is greater than the depth.
This is also consistent with the experimental phenomenon.
Therefore, on the premise that the energy distribution of the hemi-ellipsoidal shell and Gaussian laser is consistent, 
the ratio of the model's width is larger than the depth.

\begin{figure*}[htbp]
    \includegraphics[width=16cm]{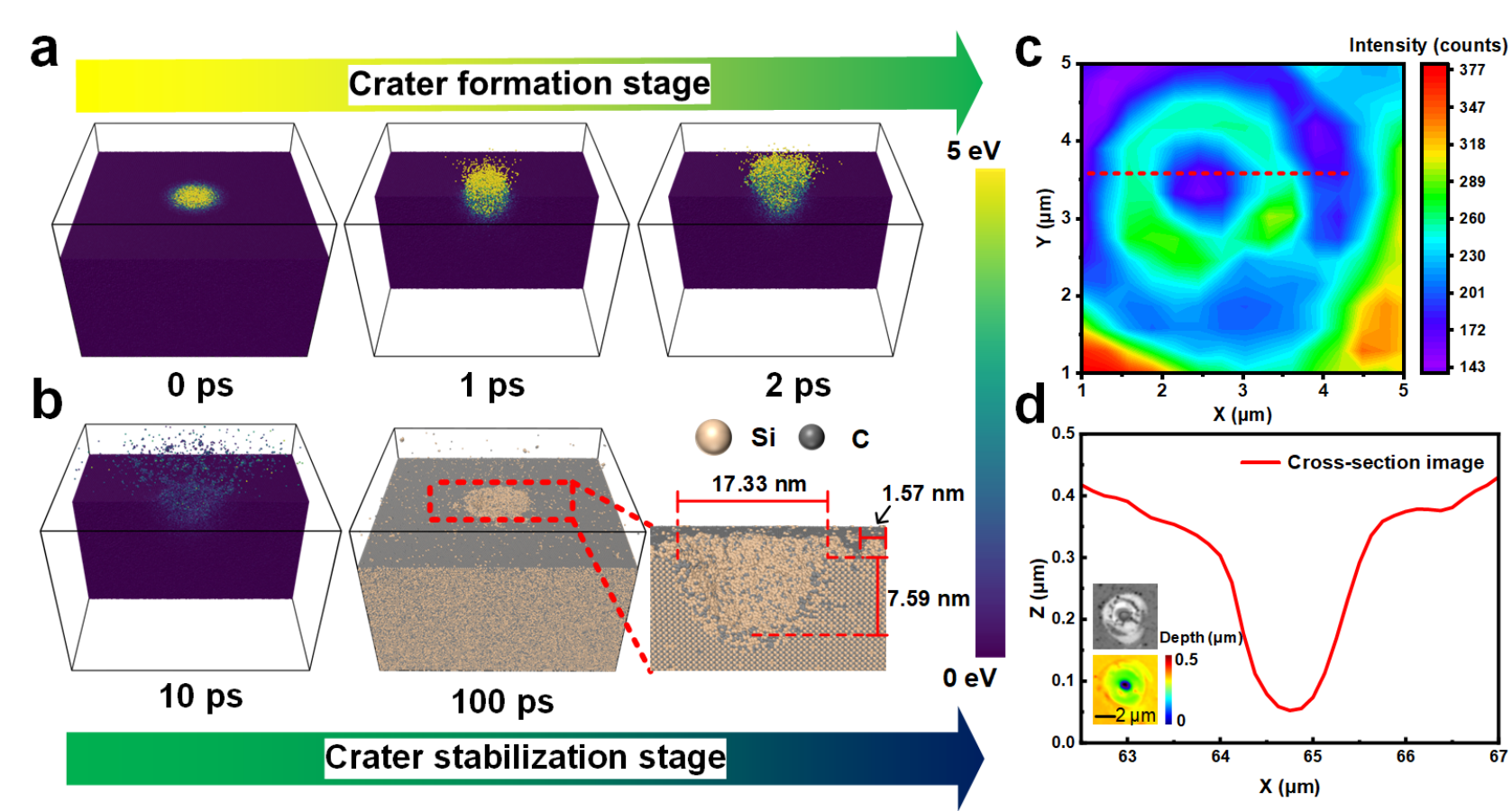}
    \caption {Particle energy distribution and morphology images using 4000 nJ laser processing. 
    (a) During the first 2 ps, 
    the process involves the formation and protrusion of knobs after energy deposition, 
    with the `crater' morphology in its formation stage. 
    (b) From 10 ps to 100 ps, the `crater' stabilizes. 
    (c) The $4H$-SiC surface after 4000 nJ processing is scanned by PL. 
    (d) The original shape and depth of $4H$-SiC surface after 4000 nJ processing scanning by Olympus laser.
    }    \label{fig:experiment_and_simulation}
\end{figure*}

In the process of energy deposition, 
Gaussian beam is divided into equal number of hemi-ellipsoidal shell layers according to the radial direction, 
and the volume is divided from the outside to the inside. 
Then, total laser energy distribution is obtained from the edge to the center of the Gaussian beam. 
As for the shell, same procedure is also applied to divide the volume from the outside to the inside, 
as shown in Figs.~\ref{fig:simulation_graph}(d) and (e).
It is necessary to ensure that the integral of the Gaussian energy distribution over the shadow region matches the energy of the corresponding filling color region of the shell.
So that the energy of each layer is determined by the total Gaussian energy corresponding to the layer divided by the volume fraction of the shell. 
Ensuring the energy of each layer has a specific value. 

After determining the energy distribution on the shell, MD simulation is used to study the laser processing on the surface of $4H$-SiC.
For the results of lower energy, the surface modification condition is gradually formed.
See more computational details in SI. 
The energy used in our test is 4000 nJ.
After the simulation, 
the density distribution of $\mathrm{V}_\mathrm{Si}$ is analyzed by 3-coordination C atoms on the surface of the material.
Figs.~\ref{fig:experiment_and_simulation}(a) and (b) reflects the details of the energy deposition of Gaussian beams which are simulated using the hemi-ellipsoidal shell. 
A large number of enriched SiC particles appeared around the processed area, 
forming a `caldera-like' uplift phenomenon. 
From Fig.~\ref{fig:experiment_and_simulation}(a), 
during the initial stage of laser energy deposition on the surface, 
a `knob-like' structure forms within 2 ps.
At this stage, particles within the processing area gain laser energy, 
transferring energy towards the surface. 
This process gradually ejects particles from the system into vacuum. 
The scattered particles within the system and the energy of the block itself are unstable, 
with a noticeable high-energy bias at the surface.
As shown in Fig.~\ref{fig:experiment_and_simulation}(b), 
the stable formation of the crater occurs between 10 ps and 100 ps. 
During this period, no new SiC particles overflow from the crater; 
instead, 
the block reaches overall thermal equilibrium, 
and particles in vacuum gradually escape and stabilize on the surface. 
By 100 ps,
\begin{figure*}[htbp]
    \includegraphics[width=16cm]{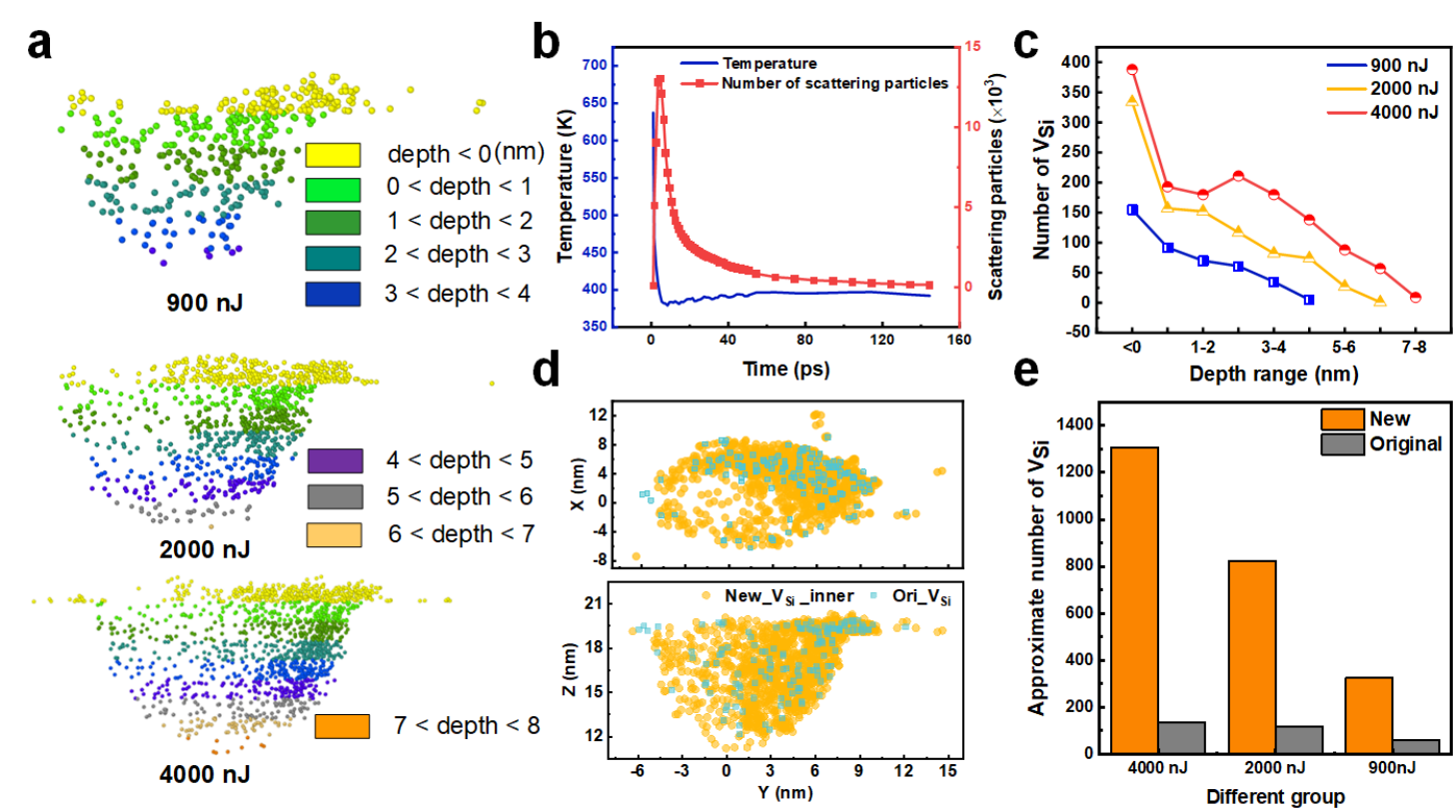}%
    \caption {Analysis the density and source of $\mathrm{V}_\mathrm{Si}$ in the system after processing at different energies.
    (a) The distribution of $\mathrm{V}_\mathrm{Si}$ after processing with 900 nJ, 2000 nJ, and 4000 nJ laser energies. Different colors indicate variations in analysis depth.
    (b) The relationship between scattering particles and temperature over time during simulated processing at 4000 nJ. 
    (c) The distribution of $\mathrm{V}_\mathrm{Si}$ at different crater depths in the stabilized system for three different processing energies.
    (d) The source of $\mathrm{V}_\mathrm{Si}$ on the surface after processing at 4000 nJ. Yellow circular dots represent $\mathrm{V}_\mathrm{Si}$ that newly formed, while blue dots represent original $\mathrm{V}_\mathrm{Si}$ structures.
    (e) The quantitative analysis of the number of $\mathrm{V}_\mathrm{Si}$ from two different sources under three different processing energies. Orange represents $\mathrm{V}_\mathrm{Si}$ that newly formed, while gray represents $\mathrm{V}_\mathrm{Si}$ that originally existed in the system.
    }
    \label{fig:source}
\end{figure*} 
detailed local enlargements of the image clearly show the `crater' stabilizing. 
At this point, 
the distribution of $\mathrm{V}_\mathrm{Si}$ on the surface is also largely fixed.
The final `crater' on the surface has a diameter of approximately 17 nm and a depth of about 7.6 nm.
SiC nanoclusters with a diameter of about 1.6 nm also exist.
The figure shows that the depth to width ratio corresponds well with the experiments.
The results are consistent with those obtained by laser processing as shown in Figs.~\ref{fig:experiment_and_simulation}(c) and (d). 
Based on the simulations and the experimental results presented by PL,
the signal of $\mathrm{V}_\mathrm{Si}$ shows an enhanced phenomenon around the processed crater, 
while the signal exhibits a significant decline within the center of the crater. 
According to microscopic images, 
the particles around the circular pit appear to accumulate, 
causing the section to bulge and the $\mathrm{V}_\mathrm{Si}$ color center density to increase. 
Therefore, the signal increases.

After measuring the width and depth of the `crater' formed by processing, 
the width is 1.625 $\upmu$m while the depth is 0.4036 $\upmu$m. 
Additional processing data are supplemented in the SI.
The results preliminarily confirm the agreement between the established model and the experimental phenomenon and the simulation results explain the observed experimental phenomenon.

In the simulation tests and experiments, 
in order to understand the source of the $\mathrm{V}_\mathrm{Si}$ generated after laser processing, 
the quantity and the source of $\mathrm{V}_\mathrm{Si}$ at different moments from the depth are analyzed.
In the phenomenon analysis shown in Fig.~\ref{fig:source}(a), 
higher energy results in deeper and wider craters, 
with $\mathrm{V}_\mathrm{Si}$ predominantly accumulating on the surface. 
Fig.~\ref{fig:source}(b) shows the scattering particles rising rapidly to 13,038 at first then gradually decreasing to 150 around 100 ps, 
and temperature stabilizes, indicating the system is stable and analyzable. 
At this stage, the distribution of $\mathrm{V}_\mathrm{Si}$ around the crater remains unchanged. 
In Fig.~\ref{fig:source}(c), 
various processing energies exhibit the characteristic of $\mathrm{V}_\mathrm{Si}$ being enriched on the surface. 
However, 
the number of $\mathrm{V}_\mathrm{Si}$ within 0-2 nm below the surface tends to be saturated as the processing energy increases, 
especially in the results of 2000 nJ and 4000 nJ processing. 
This phenomenon occurs because particles near the surface require much less energy to escape compared to those inside, 
leading to a significant migration of $\mathrm{V}_\mathrm{Si}$ to the block surface and a marked increase in surface color center structures.
The distribution of different $\mathrm{V}_\mathrm{Si}$ origins is also clearly illustrated in Fig.~\ref{fig:source}(d). 
On the surface, $\mathrm{V}_\mathrm{Si}$ are concentrated around the crater, 
within the range of $y$-values from 3 nm to 9 nm and $x$-values from -4 nm to 8 nm. 
Regardless of whether the $\mathrm{V}_\mathrm{Si}$ are originally introduced in the system or newly formed due to laser energy deposition, 
they exhibit a consistent pattern. 
The density of $\mathrm{V}_\mathrm{Si}$ is high around the crater surface, 
while the density of $\mathrm{V}_\mathrm{Si}$ is relatively low around the crater. 
For the existing color center structures in the system, 
there is a trend of decreasing color center numbers with increasing depth. 
Regarding the overall number of color centers, laser energy is more efficient in generating new color centers in the system, 
while the number of pre-existing $\mathrm{V}_\mathrm{Si}$ also increases but at a relatively slower rate.

\begin{figure}[ht!] 
    \includegraphics[width=8.6cm]{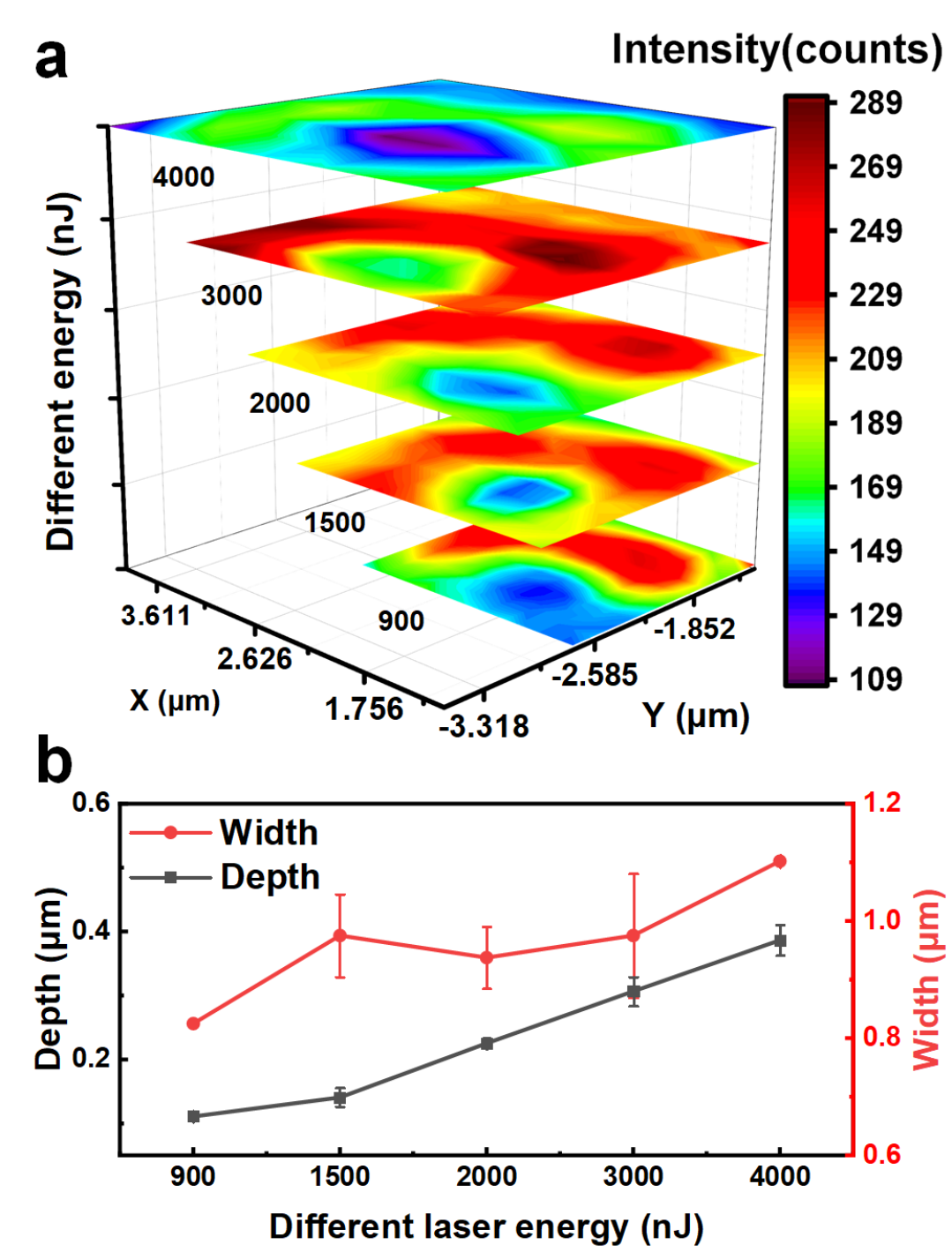}
    \caption{
        Under different pulse energy, 
        the relationship between the PL images of the $4H$-SiC surface after laser processing and size changes of the `crater's. 
        (a) The intensity information of the $\mathrm{V}_\mathrm{Si}$ signals on the surface processed with different laser energies. 
        To display more information in the images below, the images for pulse energy below 4000 nJ are scaled down. 
        (b) The relationship between the depth and diameter of the `crater's after processing with energies ranging from 900 nJ to 4000 nJ.
        }
    \label{fig:4H-SiC_different_eng}
\end{figure}

{From Fig.~\ref{fig:4H-SiC_different_eng}, 
for the laser processing with energies ranging from 900 nJ to 4000 nJ, 
the phenomenon of weaker signal in the processed areas forms a `caldera-like' structure, 
and at higher energies, a `ring-shaped' structure with enhanced signal forms is generated,
as shown in Fig.~\ref{fig:4H-SiC_different_eng}(a). 
While,
Fig.~\ref{fig:4H-SiC_different_eng} (b) explains the relationship between the crater depth and width at different laser energies: 
the crater width is always greater than the depth. 
These phenomenons are consistent with simulation results, 
which suggest that scattering particles are more likely to form a `ring-shaped' signal increasing ares on surface. 

In this work, the three-dimensional distribution of $\mathrm{V}_\mathrm{Si}$ is drawn after the simulation of 4000 nJ {\red single-pulse} processing. 
The source of color center is analyzed. Meanwhile, the overall density distribution of $\mathrm{V}_\mathrm{Si}$ in depth is also analyzed.
The physical process of 3D-MD laser energy deposition based on the geometric model of space energy distribution formed by Gaussian light and the hemi-ellipsoidal shell model is realized. 
Moreover, 
the color center density and microstructure of $4H$-SiC containing $\mathrm{V}_\mathrm{Si}$ defect are analyzed by 1030 nm laser with single pulse energy of 4000 nJ, 
pulse repetition rate of 100 kHz, which confirmed the rationality of the model. 
The python energy integration method is introduced to smooth the Gaussian energy and the energy distributed by the ellipsoidal shells, 
which provides a common reference strategy for laser processing of other semiconductor materials. 
In addition, the model also analyzes the trend of $\mathrm{V}_\mathrm{Si}$ distribution on the surface. 
This indicates that when the energy is sufficient to form significant impact craters on the surface of $4H$-SiC, 
the color center distribution is concentrated on the surface, and most of the color centers are formed newly formed $\mathrm{V}_\mathrm{Si}$ structures in the system, 
rather than the aggregation of originally existing color center structures.
This model has guiding significance to explain the color center distribution at the microscopic level.

\section*{AUTHORDECLARATIONS}
\section*{Conflict of Interest}
The authors have no conflicts to disclose.
\section*{Author Contributions}
\textbf{Mengzhi Yan:} Conceptualization (equal); Data curation (equal); Formal analysis (equal); Investigation (equal); Software (equal); Visualization (equal); Writing–original draft (leading); Writing-review \& editing (equal).
\textbf{Junlei Zhao:}
Conceptualization (leading); Data curation (leading); Formal analysis (supporting); Writing-review \& editing (equal); supervision (equal). 
\textbf{Ying Song:} Conceptualization (supporting); Data curation (supporting); Formal analysis (supporting); Writing-review \& editing (equal). 
\textbf{Bing Dong:} Resources (supporting); Writing-review \& editing (equal).
\textbf{Yifei Duan:}
Conceptualization (supporting); Data curation (supporting).
\textbf{Jianshi Wang:}
Conceptualization (supporting); Data curation (supporting).
\textbf{Qingqing Sun:}
Conceptualization (supporting); Data curation (supporting).
\textbf{Zongwei Xu:}
Project administration (leading);
Conceptualization (leading); Funding acquisition (leading); Project administration (leading); Supervision (leading);

\section*{Acknowledgments}
Project is also supported by State key laboratory of precision measuring technology and instruments (Tianjin University) under Grand pilab2203.
The study is supported by National Natural Science Foundation of China (No. 51575389, 51761135106), 2020 Mobility Programme of the Sino-German Center for Research Promotion (M-0396), and State key laboratory of precision measuring technology and instruments (Pilt1705, Pilt2107).

\section*{DATA AVAILABILITY}
The data that support the findings of this study are available from the corresponding author upon reasonable request.
Detail of energy deposition could be found on link: 
\url{https://github.com/YanMdGamma/3D_MD_laser_Model.git}         

\bibliography{ref}

\begin{thebibliography}{34}%
\makeatletter
\providecommand \@ifxundefined [1]{%
 \@ifx{#1\undefined}
}%
\providecommand \@ifnum [1]{%
 \ifnum #1\expandafter \@firstoftwo
 \else \expandafter \@secondoftwo
 \fi
}%
\providecommand \@ifx [1]{%
 \ifx #1\expandafter \@firstoftwo
 \else \expandafter \@secondoftwo
 \fi
}%
\providecommand \natexlab [1]{#1}%
\providecommand \enquote  [1]{``#1''}%
\providecommand \bibnamefont  [1]{#1}%
\providecommand \bibfnamefont [1]{#1}%
\providecommand \citenamefont [1]{#1}%
\providecommand \href@noop [0]{\@secondoftwo}%
\providecommand \href [0]{\begingroup \@sanitize@url \@href}%
\providecommand \@href[1]{\@@startlink{#1}\@@href}%
\providecommand \@@href[1]{\endgroup#1\@@endlink}%
\providecommand \@sanitize@url [0]{\catcode `\\12\catcode `\$12\catcode
  `\&12\catcode `\#12\catcode `\^12\catcode `\_12\catcode `\%12\relax}%
\providecommand \@@startlink[1]{}%
\providecommand \@@endlink[0]{}%
\providecommand \url  [0]{\begingroup\@sanitize@url \@url }%
\providecommand \@url [1]{\endgroup\@href {#1}{\urlprefix }}%
\providecommand \urlprefix  [0]{URL }%
\providecommand \Eprint [0]{\href }%
\providecommand \doibase [0]{https://doi.org/}%
\providecommand \selectlanguage [0]{\@gobble}%
\providecommand \bibinfo  [0]{\@secondoftwo}%
\providecommand \bibfield  [0]{\@secondoftwo}%
\providecommand \translation [1]{[#1]}%
\providecommand \BibitemOpen [0]{}%
\providecommand \bibitemStop [0]{}%
\providecommand \bibitemNoStop [0]{.\EOS\space}%
\providecommand \EOS [0]{\spacefactor3000\relax}%
\providecommand \BibitemShut  [1]{\csname bibitem#1\endcsname}%
\let\auto@bib@innerbib\@empty
\bibitem [{\citenamefont {Kimoto}(2019)}]{2019_property_SiC}%
  \BibitemOpen
  \bibfield  {author} {\bibinfo {author} {\bibfnamefont {T.}~\bibnamefont
  {Kimoto}}\ }(\bibinfo  {publisher} {Woodhead Publ.},\ \bibinfo {year}
  {2019})\ pp.\ \bibinfo {pages} {21--42}\BibitemShut {NoStop}%
\bibitem [{\citenamefont {Bose}\ and\ \citenamefont
  {Mazumder}(2011)}]{2011_elec_property}%
  \BibitemOpen
  \bibfield  {author} {\bibinfo {author} {\bibfnamefont {S.}~\bibnamefont
  {Bose}}\ and\ \bibinfo {author} {\bibfnamefont {S.~K.}\ \bibnamefont
  {Mazumder}},\ }\href
  {https://doi.org/https://doi.org/10.1016/j.sse.2011.03.008} {\bibfield
  {journal} {\bibinfo  {journal} {Solid-State Electron.}\ }\textbf {\bibinfo
  {volume} {62}},\ \bibinfo {pages} {5--13} (\bibinfo {year}
  {2011})}\BibitemShut {NoStop}%
\bibitem [{\citenamefont {Xiong}\ \emph {et~al.}(2022)\citenamefont {Xiong},
  \citenamefont {Mao}, \citenamefont {Wang}, \citenamefont {Liu}, \citenamefont
  {Zhang}, \citenamefont {Song}, \citenamefont {Yang},\ and\ \citenamefont
  {Pi}}]{2022_absorption_ntype}%
  \BibitemOpen
  \bibfield  {author} {\bibinfo {author} {\bibfnamefont {H.}~\bibnamefont
  {Xiong}}, \bibinfo {author} {\bibfnamefont {W.}~\bibnamefont {Mao}}, \bibinfo
  {author} {\bibfnamefont {R.}~\bibnamefont {Wang}}, \bibinfo {author}
  {\bibfnamefont {S.}~\bibnamefont {Liu}}, \bibinfo {author} {\bibfnamefont
  {N.}~\bibnamefont {Zhang}}, \bibinfo {author} {\bibfnamefont
  {L.}~\bibnamefont {Song}}, \bibinfo {author} {\bibfnamefont {D.}~\bibnamefont
  {Yang}},\ and\ \bibinfo {author} {\bibfnamefont {X.}~\bibnamefont {Pi}},\
  }\href {https://doi.org/https://doi.org/10.1016/j.mtphys.2022.100906}
  {\bibfield  {journal} {\bibinfo  {journal} {Mater. Today Phys.}\ }\textbf
  {\bibinfo {volume} {29}},\ \bibinfo {pages} {100906} (\bibinfo {year}
  {2022})}\BibitemShut {NoStop}%
\bibitem [{\citenamefont {Castelletto}\ and\ \citenamefont
  {Boretti}(2020)}]{2020_quantum_review}%
  \BibitemOpen
  \bibfield  {author} {\bibinfo {author} {\bibfnamefont {S.}~\bibnamefont
  {Castelletto}}\ and\ \bibinfo {author} {\bibfnamefont {A.}~\bibnamefont
  {Boretti}},\ }\href {https://doi.org/10.1088/2515-7647/ab77a2} {\bibfield
  {journal} {\bibinfo  {journal} {J. Phys.: Photonics}\ }\textbf {\bibinfo
  {volume} {2}},\ \bibinfo {pages} {022001} (\bibinfo {year}
  {2020})}\BibitemShut {NoStop}%
\bibitem [{\citenamefont {Lienhard}\ \emph {et~al.}(2016)\citenamefont
  {Lienhard}, \citenamefont {Schröder}, \citenamefont {Mouradian},
  \citenamefont {Dolde}, \citenamefont {Tran}, \citenamefont {Aharonovich},\
  and\ \citenamefont {Englund}}]{2016_review_quantum}%
  \BibitemOpen
  \bibfield  {author} {\bibinfo {author} {\bibfnamefont {B.}~\bibnamefont
  {Lienhard}}, \bibinfo {author} {\bibfnamefont {T.}~\bibnamefont {Schröder}},
  \bibinfo {author} {\bibfnamefont {S.}~\bibnamefont {Mouradian}}, \bibinfo
  {author} {\bibfnamefont {F.}~\bibnamefont {Dolde}}, \bibinfo {author}
  {\bibfnamefont {T.~T.}\ \bibnamefont {Tran}}, \bibinfo {author}
  {\bibfnamefont {I.}~\bibnamefont {Aharonovich}},\ and\ \bibinfo {author}
  {\bibfnamefont {D.}~\bibnamefont {Englund}},\ }\href
  {https://doi.org/10.1364/OPTICA.3.000768} {\bibfield  {journal} {\bibinfo
  {journal} {Optica}\ }\textbf {\bibinfo {volume} {3}},\ \bibinfo {pages} {768
  – 774} (\bibinfo {year} {2016})}\BibitemShut {NoStop}%
\bibitem [{\citenamefont {Childress}, \citenamefont {Walsworth},\ and\
  \citenamefont {Lukin}(2014)}]{2014_quantum}%
  \BibitemOpen
  \bibfield  {author} {\bibinfo {author} {\bibfnamefont {L.}~\bibnamefont
  {Childress}}, \bibinfo {author} {\bibfnamefont {R.}~\bibnamefont
  {Walsworth}},\ and\ \bibinfo {author} {\bibfnamefont {M.}~\bibnamefont
  {Lukin}},\ }\href {https://doi.org/10.1063/PT.3.2549} {\bibfield  {journal}
  {\bibinfo  {journal} {Phys. Today}\ }\textbf {\bibinfo {volume} {67}},\
  \bibinfo {pages} {38--43} (\bibinfo {year} {2014})}\BibitemShut {NoStop}%
\bibitem [{\citenamefont {Song}\ \emph {et~al.}(2024)\citenamefont {Song},
  \citenamefont {Xu}, \citenamefont {Rommel}, \citenamefont {Astakhov},
  \citenamefont {Hlawacek},\ and\ \citenamefont {Fang}}]{2024_4H_colorcen}%
  \BibitemOpen
  \bibfield  {author} {\bibinfo {author} {\bibfnamefont {Y.}~\bibnamefont
  {Song}}, \bibinfo {author} {\bibfnamefont {Z.}~\bibnamefont {Xu}}, \bibinfo
  {author} {\bibfnamefont {M.}~\bibnamefont {Rommel}}, \bibinfo {author}
  {\bibfnamefont {G.~V.}\ \bibnamefont {Astakhov}}, \bibinfo {author}
  {\bibfnamefont {G.}~\bibnamefont {Hlawacek}},\ and\ \bibinfo {author}
  {\bibfnamefont {F.}~\bibnamefont {Fang}},\ }\href
  {https://doi.org/https://doi.org/10.1016/j.ceramint.2023.12.096} {\bibfield
  {journal} {\bibinfo  {journal} {Ceram. Int.}\ }\textbf {\bibinfo {volume}
  {50}},\ \bibinfo {pages} {7691--7701} (\bibinfo {year} {2024})}\BibitemShut
  {NoStop}%
\bibitem [{\citenamefont {Wolfowicz}, \citenamefont {Whiteley},\ and\
  \citenamefont {Awschalom}(2018)}]{2018_elec_SiC}%
  \BibitemOpen
  \bibfield  {author} {\bibinfo {author} {\bibfnamefont {G.}~\bibnamefont
  {Wolfowicz}}, \bibinfo {author} {\bibfnamefont {S.~J.}\ \bibnamefont
  {Whiteley}},\ and\ \bibinfo {author} {\bibfnamefont {D.~D.}\ \bibnamefont
  {Awschalom}},\ }\href {https://doi.org/10.1073/pnas.1806998115} {\bibfield
  {journal} {\bibinfo  {journal} {Proc. Natl. Acad. Sci}\ }\textbf {\bibinfo
  {volume} {115}},\ \bibinfo {pages} {7879--7883} (\bibinfo {year}
  {2018})}\BibitemShut {NoStop}%
\bibitem [{\citenamefont {Wang}\ \emph {et~al.}(2023)\citenamefont {Wang},
  \citenamefont {Liu}, \citenamefont {Liu}, \citenamefont {Li}, \citenamefont
  {Cui}, \citenamefont {Zhou}, \citenamefont {Zhou}, \citenamefont {Wei},
  \citenamefont {Xu}, \citenamefont {Xu}, \citenamefont {Lin}, \citenamefont
  {Yan}, \citenamefont {He}, \citenamefont {Liu}, \citenamefont {Hao},
  \citenamefont {Li}, \citenamefont {Liu}, \citenamefont {Xu}, \citenamefont
  {Gregoryanz}, \citenamefont {Li},\ and\ \citenamefont
  {Guo}}]{2023_SiC_Magnet}%
  \BibitemOpen
  \bibfield  {author} {\bibinfo {author} {\bibfnamefont {J.-F.}\ \bibnamefont
  {Wang}}, \bibinfo {author} {\bibfnamefont {L.}~\bibnamefont {Liu}}, \bibinfo
  {author} {\bibfnamefont {X.-D.}\ \bibnamefont {Liu}}, \bibinfo {author}
  {\bibfnamefont {Q.}~\bibnamefont {Li}}, \bibinfo {author} {\bibfnamefont
  {J.-M.}\ \bibnamefont {Cui}}, \bibinfo {author} {\bibfnamefont {D.-F.}\
  \bibnamefont {Zhou}}, \bibinfo {author} {\bibfnamefont {J.-Y.}\ \bibnamefont
  {Zhou}}, \bibinfo {author} {\bibfnamefont {Y.}~\bibnamefont {Wei}}, \bibinfo
  {author} {\bibfnamefont {H.-A.}\ \bibnamefont {Xu}}, \bibinfo {author}
  {\bibfnamefont {W.}~\bibnamefont {Xu}}, \bibinfo {author} {\bibfnamefont
  {W.-X.}\ \bibnamefont {Lin}}, \bibinfo {author} {\bibfnamefont {J.-W.}\
  \bibnamefont {Yan}}, \bibinfo {author} {\bibfnamefont {Z.-X.}\ \bibnamefont
  {He}}, \bibinfo {author} {\bibfnamefont {Z.-H.}\ \bibnamefont {Liu}},
  \bibinfo {author} {\bibfnamefont {Z.-H.}\ \bibnamefont {Hao}}, \bibinfo
  {author} {\bibfnamefont {H.-O.}\ \bibnamefont {Li}}, \bibinfo {author}
  {\bibfnamefont {W.}~\bibnamefont {Liu}}, \bibinfo {author} {\bibfnamefont
  {J.-S.}\ \bibnamefont {Xu}}, \bibinfo {author} {\bibfnamefont
  {E.}~\bibnamefont {Gregoryanz}}, \bibinfo {author} {\bibfnamefont {C.-F.}\
  \bibnamefont {Li}},\ and\ \bibinfo {author} {\bibfnamefont {G.-C.}\
  \bibnamefont {Guo}},\ }\href {https://doi.org/10.1038/s41563-023-01477-5}
  {\bibfield  {journal} {\bibinfo  {journal} {Nat. Mater.}\ }\textbf {\bibinfo
  {volume} {22}},\ \bibinfo {pages} {489--494} (\bibinfo {year}
  {2023})}\BibitemShut {NoStop}%
\bibitem [{\citenamefont {Anisimov}\ \emph {et~al.}(2016)\citenamefont
  {Anisimov}, \citenamefont {Simin}, \citenamefont {Soltamov}, \citenamefont
  {Lebedev}, \citenamefont {Baranov}, \citenamefont {Astakhov},\ and\
  \citenamefont {Dyakonov}}]{2016_SiC_Therm}%
  \BibitemOpen
  \bibfield  {author} {\bibinfo {author} {\bibfnamefont {A.~N.}\ \bibnamefont
  {Anisimov}}, \bibinfo {author} {\bibfnamefont {D.}~\bibnamefont {Simin}},
  \bibinfo {author} {\bibfnamefont {V.~A.}\ \bibnamefont {Soltamov}}, \bibinfo
  {author} {\bibfnamefont {S.~P.}\ \bibnamefont {Lebedev}}, \bibinfo {author}
  {\bibfnamefont {P.~G.}\ \bibnamefont {Baranov}}, \bibinfo {author}
  {\bibfnamefont {G.~V.}\ \bibnamefont {Astakhov}},\ and\ \bibinfo {author}
  {\bibfnamefont {V.}~\bibnamefont {Dyakonov}},\ }\href
  {https://doi.org/10.1038/srep33301} {\bibfield  {journal} {\bibinfo
  {journal} {Sci. Rep.}\ }\textbf {\bibinfo {volume} {6}},\ \bibinfo {pages}
  {33301} (\bibinfo {year} {2016})}\BibitemShut {NoStop}%
\bibitem [{\citenamefont {Zhou}\ \emph {et~al.}(2017)\citenamefont {Zhou},
  \citenamefont {Wang}, \citenamefont {Zhang}, \citenamefont {Li},
  \citenamefont {Cai},\ and\ \citenamefont {Gao}}]{2017_SiC_Therm}%
  \BibitemOpen
  \bibfield  {author} {\bibinfo {author} {\bibfnamefont {Y.}~\bibnamefont
  {Zhou}}, \bibinfo {author} {\bibfnamefont {J.}~\bibnamefont {Wang}}, \bibinfo
  {author} {\bibfnamefont {X.}~\bibnamefont {Zhang}}, \bibinfo {author}
  {\bibfnamefont {K.}~\bibnamefont {Li}}, \bibinfo {author} {\bibfnamefont
  {J.}~\bibnamefont {Cai}},\ and\ \bibinfo {author} {\bibfnamefont
  {W.}~\bibnamefont {Gao}},\ }\href
  {https://doi.org/10.1103/PhysRevApplied.8.044015} {\bibfield  {journal}
  {\bibinfo  {journal} {Phys. Rev. Appl.}\ }\textbf {\bibinfo {volume} {8}},\
  \bibinfo {pages} {044015} (\bibinfo {year} {2017})}\BibitemShut {NoStop}%
\bibitem [{\citenamefont {Koehl}\ \emph {et~al.}(2011)\citenamefont {Koehl},
  \citenamefont {Buckley}, \citenamefont {Heremans}, \citenamefont {Calusine},\
  and\ \citenamefont {Awschalom}}]{2011_colorcen}%
  \BibitemOpen
  \bibfield  {author} {\bibinfo {author} {\bibfnamefont {W.~F.}\ \bibnamefont
  {Koehl}}, \bibinfo {author} {\bibfnamefont {B.~B.}\ \bibnamefont {Buckley}},
  \bibinfo {author} {\bibfnamefont {F.~J.}\ \bibnamefont {Heremans}}, \bibinfo
  {author} {\bibfnamefont {G.}~\bibnamefont {Calusine}},\ and\ \bibinfo
  {author} {\bibfnamefont {D.~D.}\ \bibnamefont {Awschalom}},\ }\href
  {https://doi.org/10.1038/nature10562} {\bibfield  {journal} {\bibinfo
  {journal} {Nature}\ }\textbf {\bibinfo {volume} {479}},\ \bibinfo {pages}
  {84--87} (\bibinfo {year} {2011})}\BibitemShut {NoStop}%
\bibitem [{\citenamefont {Castelletto}\ \emph {et~al.}(2014)\citenamefont
  {Castelletto}, \citenamefont {Johnson}, \citenamefont {Ivády}, \citenamefont
  {Stavrias}, \citenamefont {Umeda}, \citenamefont {Gali},\ and\ \citenamefont
  {Ohshima}}]{2014_ion_im}%
  \BibitemOpen
  \bibfield  {author} {\bibinfo {author} {\bibfnamefont {S.}~\bibnamefont
  {Castelletto}}, \bibinfo {author} {\bibfnamefont {B.~C.}\ \bibnamefont
  {Johnson}}, \bibinfo {author} {\bibfnamefont {V.}~\bibnamefont {Ivády}},
  \bibinfo {author} {\bibfnamefont {N.}~\bibnamefont {Stavrias}}, \bibinfo
  {author} {\bibfnamefont {T.}~\bibnamefont {Umeda}}, \bibinfo {author}
  {\bibfnamefont {A.}~\bibnamefont {Gali}},\ and\ \bibinfo {author}
  {\bibfnamefont {T.}~\bibnamefont {Ohshima}},\ }\href
  {https://doi.org/10.1038/nmat3806} {\bibfield  {journal} {\bibinfo  {journal}
  {Nat. Mater.}\ }\textbf {\bibinfo {volume} {13}},\ \bibinfo {pages}
  {151--156} (\bibinfo {year} {2014})}\BibitemShut {NoStop}%
\bibitem [{\citenamefont {Kimoto}\ and\ \citenamefont
  {Cooper}(2014)}]{2014_chem_colorcen}%
  \BibitemOpen
  \bibfield  {author} {\bibinfo {author} {\bibfnamefont {T.}~\bibnamefont
  {Kimoto}}\ and\ \bibinfo {author} {\bibfnamefont {J.~A.}\ \bibnamefont
  {Cooper}},\ }\href@noop {} {}\ (\bibinfo  {publisher} {John Wiley \& Sons},\
  \bibinfo {year} {2014})\BibitemShut {NoStop}%
\bibitem [{\citenamefont {von Bardeleben}\ \emph {et~al.}(2015)\citenamefont
  {von Bardeleben}, \citenamefont {Cantin}, \citenamefont {Rauls},\ and\
  \citenamefont {Gerstmann}}]{2015_ther_colorcen}%
  \BibitemOpen
  \bibfield  {author} {\bibinfo {author} {\bibfnamefont {H.~J.}\ \bibnamefont
  {von Bardeleben}}, \bibinfo {author} {\bibfnamefont {J.~L.}\ \bibnamefont
  {Cantin}}, \bibinfo {author} {\bibfnamefont {E.}~\bibnamefont {Rauls}},\ and\
  \bibinfo {author} {\bibfnamefont {U.}~\bibnamefont {Gerstmann}},\ }\href
  {https://doi.org/10.1103/PhysRevB.92.064104} {\bibfield  {journal} {\bibinfo
  {journal} {Phys. Rev. B}\ }\textbf {\bibinfo {volume} {92}},\ \bibinfo
  {pages} {064104} (\bibinfo {year} {2015})}\BibitemShut {NoStop}%
\bibitem [{\citenamefont {Gao}\ \emph {et~al.}(2022)\citenamefont {Gao},
  \citenamefont {Duan}, \citenamefont {Tian}, \citenamefont {Zhang},
  \citenamefont {Chen}, \citenamefont {Gao},\ and\ \citenamefont
  {Sun}}]{2022_laser_SiC}%
  \BibitemOpen
  \bibfield  {author} {\bibinfo {author} {\bibfnamefont {S.}~\bibnamefont
  {Gao}}, \bibinfo {author} {\bibfnamefont {Y.-Z.}\ \bibnamefont {Duan}},
  \bibinfo {author} {\bibfnamefont {Z.-N.}\ \bibnamefont {Tian}}, \bibinfo
  {author} {\bibfnamefont {Y.-L.}\ \bibnamefont {Zhang}}, \bibinfo {author}
  {\bibfnamefont {Q.-D.}\ \bibnamefont {Chen}}, \bibinfo {author}
  {\bibfnamefont {B.-R.}\ \bibnamefont {Gao}},\ and\ \bibinfo {author}
  {\bibfnamefont {H.-B.}\ \bibnamefont {Sun}},\ }\href
  {https://doi.org/https://doi.org/10.1016/j.optlastec.2021.107527} {\bibfield
  {journal} {\bibinfo  {journal} {Opt. Laser Technol.}\ }\textbf {\bibinfo
  {volume} {146}},\ \bibinfo {pages} {107527} (\bibinfo {year}
  {2022})}\BibitemShut {NoStop}%
\bibitem [{\citenamefont {Rehman}\ and\ \citenamefont
  {Janulewicz}(2016)}]{2016_SiC_laser_app}%
  \BibitemOpen
  \bibfield  {author} {\bibinfo {author} {\bibfnamefont {Z.}~\bibnamefont
  {Rehman}}\ and\ \bibinfo {author} {\bibfnamefont {K.}~\bibnamefont
  {Janulewicz}},\ }\href
  {https://doi.org/https://doi.org/10.1016/j.apsusc.2016.05.041} {\bibfield
  {journal} {\bibinfo  {journal} {Appl. Surf. Sci.}\ }\textbf {\bibinfo
  {volume} {385}},\ \bibinfo {pages} {1--8} (\bibinfo {year}
  {2016})}\BibitemShut {NoStop}%
\bibitem [{\citenamefont {Huang}\ \emph {et~al.}(2023)\citenamefont {Huang},
  \citenamefont {Zhou}, \citenamefont {Li},\ and\ \citenamefont
  {Zhu}}]{2023_4H_laserMD}%
  \BibitemOpen
  \bibfield  {author} {\bibinfo {author} {\bibfnamefont {Y.}~\bibnamefont
  {Huang}}, \bibinfo {author} {\bibfnamefont {Y.}~\bibnamefont {Zhou}},
  \bibinfo {author} {\bibfnamefont {J.}~\bibnamefont {Li}},\ and\ \bibinfo
  {author} {\bibfnamefont {F.}~\bibnamefont {Zhu}},\ }\href
  {https://doi.org/https://doi.org/10.1016/j.apsusc.2023.156436} {\bibfield
  {journal} {\bibinfo  {journal} {Appl. Surf. Sci.}\ }\textbf {\bibinfo
  {volume} {615}},\ \bibinfo {pages} {156436} (\bibinfo {year}
  {2023})}\BibitemShut {NoStop}%
\bibitem [{\citenamefont {Dong}\ \emph {et~al.}(2023)\citenamefont {Dong},
  \citenamefont {Xu}, \citenamefont {Shi}, \citenamefont {Zhang}, \citenamefont
  {Zhang}, \citenamefont {Hua}, \citenamefont {Zhao},\ and\ \citenamefont
  {Wang}}]{2023_4H_laser}%
  \BibitemOpen
  \bibfield  {author} {\bibinfo {author} {\bibfnamefont {B.}~\bibnamefont
  {Dong}}, \bibinfo {author} {\bibfnamefont {Z.}~\bibnamefont {Xu}}, \bibinfo
  {author} {\bibfnamefont {C.}~\bibnamefont {Shi}}, \bibinfo {author}
  {\bibfnamefont {K.}~\bibnamefont {Zhang}}, \bibinfo {author} {\bibfnamefont
  {Y.}~\bibnamefont {Zhang}}, \bibinfo {author} {\bibfnamefont
  {R.}~\bibnamefont {Hua}}, \bibinfo {author} {\bibfnamefont {W.}~\bibnamefont
  {Zhao}},\ and\ \bibinfo {author} {\bibfnamefont {J.}~\bibnamefont {Wang}},\
  }\href {https://doi.org/https://doi.org/10.1016/j.optlastec.2023.109338}
  {\bibfield  {journal} {\bibinfo  {journal} {Opt. Laser Technol.}\ }\textbf
  {\bibinfo {volume} {163}},\ \bibinfo {pages} {109338} (\bibinfo {year}
  {2023})}\BibitemShut {NoStop}%
\bibitem [{\citenamefont {Zargaleh}\ \emph {et~al.}(2016)\citenamefont
  {Zargaleh}, \citenamefont {Eble}, \citenamefont {Hameau}, \citenamefont
  {Cantin}, \citenamefont {Legrand}, \citenamefont {Bernard}, \citenamefont
  {Margaillan}, \citenamefont {Lauret}, \citenamefont {Roch}, \citenamefont
  {von Bardeleben}, \citenamefont {Rauls}, \citenamefont {Gerstmann},\ and\
  \citenamefont {Treussart}}]{2016_colorcen}%
  \BibitemOpen
  \bibfield  {author} {\bibinfo {author} {\bibfnamefont {S.~A.}\ \bibnamefont
  {Zargaleh}}, \bibinfo {author} {\bibfnamefont {B.}~\bibnamefont {Eble}},
  \bibinfo {author} {\bibfnamefont {S.}~\bibnamefont {Hameau}}, \bibinfo
  {author} {\bibfnamefont {J.-L.}\ \bibnamefont {Cantin}}, \bibinfo {author}
  {\bibfnamefont {L.}~\bibnamefont {Legrand}}, \bibinfo {author} {\bibfnamefont
  {M.}~\bibnamefont {Bernard}}, \bibinfo {author} {\bibfnamefont
  {F.}~\bibnamefont {Margaillan}}, \bibinfo {author} {\bibfnamefont {J.-S.}\
  \bibnamefont {Lauret}}, \bibinfo {author} {\bibfnamefont {J.-F.}\
  \bibnamefont {Roch}}, \bibinfo {author} {\bibfnamefont {H.~J.}\ \bibnamefont
  {von Bardeleben}}, \bibinfo {author} {\bibfnamefont {E.}~\bibnamefont
  {Rauls}}, \bibinfo {author} {\bibfnamefont {U.}~\bibnamefont {Gerstmann}},\
  and\ \bibinfo {author} {\bibfnamefont {F.}~\bibnamefont {Treussart}},\ }\href
  {https://doi.org/10.1103/PhysRevB.94.060102} {\bibfield  {journal} {\bibinfo
  {journal} {Phys. Rev. B}\ }\textbf {\bibinfo {volume} {94}},\ \bibinfo
  {pages} {060102} (\bibinfo {year} {2016})}\BibitemShut {NoStop}%
\bibitem [{\citenamefont {Adelmann}\ and\ \citenamefont
  {Hellmann}(2016)}]{2016_SiC_COMSOL}%
  \BibitemOpen
  \bibfield  {author} {\bibinfo {author} {\bibfnamefont {B.}~\bibnamefont
  {Adelmann}}\ and\ \bibinfo {author} {\bibfnamefont {R.}~\bibnamefont
  {Hellmann}},\ }\href {https://doi.org/10.1007/s00339-016-0173-x} {\bibfield
  {journal} {\bibinfo  {journal} {Appl. Phys. A}\ }\textbf {\bibinfo {volume}
  {122}},\ \bibinfo {pages} {642} (\bibinfo {year} {2016})}\BibitemShut
  {NoStop}%
\bibitem [{\citenamefont {Ravichandran}\ \emph {et~al.}(2021)\citenamefont
  {Ravichandran}, \citenamefont {Mehta}, \citenamefont {Woodworth},\ and\
  \citenamefont {Lawson}}]{2024_SiC_TTMMD}%
  \BibitemOpen
  \bibfield  {author} {\bibinfo {author} {\bibfnamefont {A.}~\bibnamefont
  {Ravichandran}}, \bibinfo {author} {\bibfnamefont {M.}~\bibnamefont {Mehta}},
  \bibinfo {author} {\bibfnamefont {A.~A.}\ \bibnamefont {Woodworth}},\ and\
  \bibinfo {author} {\bibfnamefont {J.~W.}\ \bibnamefont {Lawson}},\ }\href
  {https://doi.org/10.1063/5.0045766} {\bibfield  {journal} {\bibinfo
  {journal} {J. Appl. Phys.}\ }\textbf {\bibinfo {volume} {129}},\ \bibinfo
  {pages} {215304} (\bibinfo {year} {2021})}\BibitemShut {NoStop}%
\bibitem [{\citenamefont {An}\ \emph {et~al.}(2023)\citenamefont {An},
  \citenamefont {Wang}, \citenamefont {Cui},\ and\ \citenamefont
  {Fang}}]{2023_SiC_TTMMD_An}%
  \BibitemOpen
  \bibfield  {author} {\bibinfo {author} {\bibfnamefont {H.}~\bibnamefont
  {An}}, \bibinfo {author} {\bibfnamefont {J.}~\bibnamefont {Wang}}, \bibinfo
  {author} {\bibfnamefont {H.}~\bibnamefont {Cui}},\ and\ \bibinfo {author}
  {\bibfnamefont {F.}~\bibnamefont {Fang}},\ }\href
  {https://doi.org/10.1364/OE.487761} {\bibfield  {journal} {\bibinfo
  {journal} {Opt. Express}\ }\textbf {\bibinfo {volume} {31}},\ \bibinfo
  {pages} {15438--15448} (\bibinfo {year} {2023})}\BibitemShut {NoStop}%
\bibitem [{\citenamefont {Zhang}\ \emph {et~al.}(2022)\citenamefont {Zhang},
  \citenamefont {Xu}, \citenamefont {Wang}, \citenamefont {Zhang},\ and\
  \citenamefont {Dong}}]{2022_Cater_laser}%
  \BibitemOpen
  \bibfield  {author} {\bibinfo {author} {\bibfnamefont {K.}~\bibnamefont
  {Zhang}}, \bibinfo {author} {\bibfnamefont {Z.}~\bibnamefont {Xu}}, \bibinfo
  {author} {\bibfnamefont {H.}~\bibnamefont {Wang}}, \bibinfo {author}
  {\bibfnamefont {S.}~\bibnamefont {Zhang}},\ and\ \bibinfo {author}
  {\bibfnamefont {B.}~\bibnamefont {Dong}},\ }\href
  {https://doi.org/https://doi.org/10.1016/j.ceramint.2022.06.061} {\bibfield
  {journal} {\bibinfo  {journal} {Ceram. Int.}\ }\textbf {\bibinfo {volume}
  {48}},\ \bibinfo {pages} {27650--27657} (\bibinfo {year} {2022})}\BibitemShut
  {NoStop}%
\bibitem [{\citenamefont {Wang}\ \emph {et~al.}(2013)\citenamefont {Wang},
  \citenamefont {Zhan}, \citenamefont {Wang}, \citenamefont {Xuan},
  \citenamefont {Zhang}, \citenamefont {Liu}, \citenamefont {Xu}, \citenamefont
  {Liu}, \citenamefont {Wei},\ and\ \citenamefont {Chen}}]{2013_4H_reflective}%
  \BibitemOpen
  \bibfield  {author} {\bibinfo {author} {\bibfnamefont {S.}~\bibnamefont
  {Wang}}, \bibinfo {author} {\bibfnamefont {M.}~\bibnamefont {Zhan}}, \bibinfo
  {author} {\bibfnamefont {G.}~\bibnamefont {Wang}}, \bibinfo {author}
  {\bibfnamefont {H.}~\bibnamefont {Xuan}}, \bibinfo {author} {\bibfnamefont
  {W.}~\bibnamefont {Zhang}}, \bibinfo {author} {\bibfnamefont
  {C.}~\bibnamefont {Liu}}, \bibinfo {author} {\bibfnamefont {C.}~\bibnamefont
  {Xu}}, \bibinfo {author} {\bibfnamefont {Y.}~\bibnamefont {Liu}}, \bibinfo
  {author} {\bibfnamefont {Z.}~\bibnamefont {Wei}},\ and\ \bibinfo {author}
  {\bibfnamefont {X.}~\bibnamefont {Chen}},\ }\href
  {https://doi.org/https://doi.org/10.1002/lpor.201300068} {\bibfield
  {journal} {\bibinfo  {journal} {Laser Photonics Rev.}\ }\textbf {\bibinfo
  {volume} {7}},\ \bibinfo {pages} {831--838} (\bibinfo {year}
  {2013})}\BibitemShut {NoStop}%
\bibitem [{\citenamefont {Polyanskiy}(2024)}]{2024_datSiC}%
  \BibitemOpen
  \bibfield  {author} {\bibinfo {author} {\bibfnamefont {M.~N.}\ \bibnamefont
  {Polyanskiy}},\ }\href {https://doi.org/10.1038/s41597-023-02898-2}
  {\bibfield  {journal} {\bibinfo  {journal} {Sci. Data}\ }\textbf {\bibinfo
  {volume} {11}},\ \bibinfo {pages} {94} (\bibinfo {year} {2024})}\BibitemShut
  {NoStop}%
\bibitem [{\citenamefont {Zhang}, \citenamefont {Zhang},\ and\ \citenamefont
  {Luby}(2007)}]{2007_reflect}%
  \BibitemOpen
  \bibfield  {author} {\bibinfo {author} {\bibfnamefont {Z.~M.}\ \bibnamefont
  {Zhang}}, \bibinfo {author} {\bibfnamefont {Z.~M.}\ \bibnamefont {Zhang}},\
  and\ \bibinfo {author} {\bibnamefont {Luby}},\ }\href@noop {} {}Vol.\
  \bibinfo {volume} {410}\ (\bibinfo  {publisher} {Springer},\ \bibinfo {year}
  {2007})\BibitemShut {NoStop}%
\bibitem [{\citenamefont {Wang}, \citenamefont {Liu},\ and\ \citenamefont
  {Zhang}(2013)}]{2013_Refection}%
  \BibitemOpen
  \bibfield  {author} {\bibinfo {author} {\bibfnamefont {H.}~\bibnamefont
  {Wang}}, \bibinfo {author} {\bibfnamefont {X.}~\bibnamefont {Liu}},\ and\
  \bibinfo {author} {\bibfnamefont {Z.~M.}\ \bibnamefont {Zhang}},\ }\href
  {https://doi.org/10.1007/s10765-013-1414-2} {\bibfield  {journal} {\bibinfo
  {journal} {Int. J. Thermophys.}\ }\textbf {\bibinfo {volume} {34}},\ \bibinfo
  {pages} {213--225} (\bibinfo {year} {2013})}\BibitemShut {NoStop}%
\bibitem [{\citenamefont {Xu}\ \emph {et~al.}(2018)\citenamefont {Xu},
  \citenamefont {He}, \citenamefont {Song}, \citenamefont {Fu}, \citenamefont
  {Rommel}, \citenamefont {Luo}, \citenamefont {Hartmaier}, \citenamefont
  {Zhang},\ and\ \citenamefont {Fang}}]{2018_review_absorption}%
  \BibitemOpen
  \bibfield  {author} {\bibinfo {author} {\bibfnamefont {Z.}~\bibnamefont
  {Xu}}, \bibinfo {author} {\bibfnamefont {Z.}~\bibnamefont {He}}, \bibinfo
  {author} {\bibfnamefont {Y.}~\bibnamefont {Song}}, \bibinfo {author}
  {\bibfnamefont {X.}~\bibnamefont {Fu}}, \bibinfo {author} {\bibfnamefont
  {M.}~\bibnamefont {Rommel}}, \bibinfo {author} {\bibfnamefont
  {X.}~\bibnamefont {Luo}}, \bibinfo {author} {\bibfnamefont {A.}~\bibnamefont
  {Hartmaier}}, \bibinfo {author} {\bibfnamefont {J.}~\bibnamefont {Zhang}},\
  and\ \bibinfo {author} {\bibfnamefont {F.}~\bibnamefont {Fang}},\ }\href
  {https://doi.org/10.3390/mi9070361} {\bibfield  {journal} {\bibinfo
  {journal} {Micromachines}\ }\textbf {\bibinfo {volume} {9}} (\bibinfo {year}
  {2018}),\ 10.3390/mi9070361}\BibitemShut {NoStop}%
\bibitem [{\citenamefont {Derst}\ \emph {et~al.}(1989)\citenamefont {Derst},
  \citenamefont {Wilbertz}, \citenamefont {Bhatia}, \citenamefont
  {Krätschmer},\ and\ \citenamefont {Kalbitzer}}]{1989_ori}%
  \BibitemOpen
  \bibfield  {author} {\bibinfo {author} {\bibfnamefont {G.}~\bibnamefont
  {Derst}}, \bibinfo {author} {\bibfnamefont {C.}~\bibnamefont {Wilbertz}},
  \bibinfo {author} {\bibfnamefont {K.~L.}\ \bibnamefont {Bhatia}}, \bibinfo
  {author} {\bibfnamefont {W.}~\bibnamefont {Krätschmer}},\ and\ \bibinfo
  {author} {\bibfnamefont {S.}~\bibnamefont {Kalbitzer}},\ }\href
  {https://doi.org/10.1063/1.101271} {\bibfield  {journal} {\bibinfo  {journal}
  {Appl. Phys. Lett.}\ }\textbf {\bibinfo {volume} {54}},\ \bibinfo {pages}
  {1722--1724} (\bibinfo {year} {1989})}\BibitemShut {NoStop}%
\bibitem [{\citenamefont {Weingärtner}\ \emph {et~al.}(2001)\citenamefont
  {Weingärtner}, \citenamefont {Bickermann}, \citenamefont {Bushevoy},
  \citenamefont {Hofmann}, \citenamefont {Rasp}, \citenamefont {Straubinger},
  \citenamefont {Wellmann},\ and\ \citenamefont {Winnacker}}]{2001_absorb_SiC}%
  \BibitemOpen
  \bibfield  {author} {\bibinfo {author} {\bibfnamefont {R.}~\bibnamefont
  {Weingärtner}}, \bibinfo {author} {\bibfnamefont {M.}~\bibnamefont
  {Bickermann}}, \bibinfo {author} {\bibfnamefont {S.}~\bibnamefont
  {Bushevoy}}, \bibinfo {author} {\bibfnamefont {D.}~\bibnamefont {Hofmann}},
  \bibinfo {author} {\bibfnamefont {M.}~\bibnamefont {Rasp}}, \bibinfo {author}
  {\bibfnamefont {T.}~\bibnamefont {Straubinger}}, \bibinfo {author}
  {\bibfnamefont {P.}~\bibnamefont {Wellmann}},\ and\ \bibinfo {author}
  {\bibfnamefont {A.}~\bibnamefont {Winnacker}},\ }\href
  {https://doi.org/https://doi.org/10.1016/S0921-5107(00)00599-7} {\bibfield
  {journal} {\bibinfo  {journal} {Mater. Sci. Eng.: B}\ }\textbf {\bibinfo
  {volume} {80}},\ \bibinfo {pages} {357--361} (\bibinfo {year}
  {2001})}\BibitemShut {NoStop}%
\bibitem [{\citenamefont {Hecht}(2002)}]{2002_optics_ART}%
  \BibitemOpen
  \bibfield  {author} {\bibinfo {author} {\bibfnamefont {E.}~\bibnamefont
  {Hecht}},\ }\href@noop {} {}\ (\bibinfo  {publisher} {Addison-Wesley},\
  \bibinfo {year} {2002})\BibitemShut {NoStop}%
\bibitem [{\citenamefont {Angeloni}\ \emph {et~al.}(2024)\citenamefont
  {Angeloni}, \citenamefont {Shan}, \citenamefont {Leach},\ and\ \citenamefont
  {Schroeder}}]{2024_coefficient_opt}%
  \BibitemOpen
  \bibfield  {author} {\bibinfo {author} {\bibfnamefont {L.~A.}\ \bibnamefont
  {Angeloni}}, \bibinfo {author} {\bibfnamefont {I.-J.}\ \bibnamefont {Shan}},
  \bibinfo {author} {\bibfnamefont {J.~H.}\ \bibnamefont {Leach}},\ and\
  \bibinfo {author} {\bibfnamefont {W.~A.}\ \bibnamefont {Schroeder}},\ }\href
  {https://doi.org/10.1063/5.0213263} {\bibfield  {journal} {\bibinfo
  {journal} {Appl. Phys. Lett.}\ }\textbf {\bibinfo {volume} {124}},\ \bibinfo
  {pages} {252104} (\bibinfo {year} {2024})}\BibitemShut {NoStop}%
\bibitem [{\citenamefont {Hazzan}, \citenamefont {Pacella},\ and\ \citenamefont
  {See}(2021)}]{2021_fslaser_theory}%
  \BibitemOpen
  \bibfield  {author} {\bibinfo {author} {\bibfnamefont {K.~E.}\ \bibnamefont
  {Hazzan}}, \bibinfo {author} {\bibfnamefont {M.}~\bibnamefont {Pacella}},\
  and\ \bibinfo {author} {\bibfnamefont {T.~L.}\ \bibnamefont {See}},\ }\href
  {https://doi.org/10.3390/mi12080895} {\bibfield  {journal} {\bibinfo
  {journal} {Micromachines}\ }\textbf {\bibinfo {volume} {12}} (\bibinfo {year}
  {2021}),\ 10.3390/mi12080895}\BibitemShut {NoStop}%
\end{thebibliography}%
\end{document}